# Dynamical Search for Substructures in Galaxy Clusters

## A Hierarchical Clustering Method

**Arturo Serna**[1] **and Daniel Gerbal**[2]

[1] LAEC, Observatoire de Paris, Université Denis Diderot, CNRS (UA 173), F-92195 Meudon Cedex, France, serna@gin.obspm.fr
[2] Institut d'Astrophysique de Paris, CNRS, Université Pierre et Marie Curie, 98bis Bd Arago, F-75014 Paris, France, gerbal@iap.fr

September 15, 1995

**Abstract.** We propose a new hierarchical method which uses dynamical arguments to find and describe substructures in galaxy clusters. This method (hereafter h–method or h–analysis) uses a hierarchical clustering analysis to determine the relationship between galaxies according to their relative binding energies. We have tested from N-body simulations, the two following features of the proposed method:

1. It extracts subgroups which are much more stable during the cluster evolution than those given by other techniques.
2. There exists a reasonable similarity between the structures found when only the coordinates $(x, y, v_z)$ provided by "observations" are considered, and those found by using the six phase–space coordinates

We have applied this method to two Abell clusters: ABCG151 and ABCG2670. Our results imply that
ABCG151 is separated into two clusters, one of them is again divided into two subclusters.
ABCG2670 has no subclustering. Our method allows however to extract the most bound galaxies in its dynamical core.

## 1. Introduction

We address here the question of how to extract subclusters – or substructures – from a galaxy cluster?

The existence of substructure in galaxy clusters was first noticed from optical data by Baier (1977) and, in an extensive study, by Geller and Beers (1982). Since the work by Forman & Jones (1981) it has also been detected from X-ray observations. Such a small-scale structure may provide us with very valuable information on the internal cluster dynamics, and/or on cosmological scenarios of cluster formation. Several works (in the hierarchical formation scenario) show in fact that the existence of substructure in a galaxy cluster could be related to intrinsic parameters as its evolutionary status, the internal amount and distribution of dark matter, etc. (see, e.g., Serna, Alimi & Schol 1994, and references therein).

Moreover, the interpretation of subclusters as recent merger events might explain some of the cD galaxy properties, as their orientation with respect to the environment (West 1994a), and the peculiar velocity distribution in rich clusters (Merritt 1985; Tremaine 1990). Still in the hierarchical formation framework, it has been suggested that the frequency of substructure in galaxy clusters may constrain the density parameter $\Omega$ and the spectrum of primordial density fluctuations (Richstone, Loeb & Turner 1992; Lacey & Cole 1993; Kauffmann & White 1993).

The development of methods to detect and identify subclusters is therefore necessary to link observations with theoretical predictions and models. A huge number of papers has been devoted to this subject and several substructure detection methods have been proposed in the past. An exhaustive list of these methods can be found, for instance, in Bird (1995) or West (1994b). We will just mention here some techniques not referred to by these authors, as the use of a correlation function analysis proposed by Salvador-Sole et al (1993), or the wavelet analysis on X-ray images by Slezak et al. (1994) and Grebenev et al. (1995).

We could roughly summarize most of the substructure detection methods as essentially based on searching for overprobable coincidences and correlations in the space of positions and/or velocities. Most of them just inform that substructure is probably present (or not) in a galaxy cluster, or find those regions in the sky clearly dominated by subclusters.

We consider that the problem of identifying substructures in a galaxy cluster cannot be separated of questions like: how can substructures be defined from a dynamical point of view?, are they *transient* or *quasi-permanent* phenomena?, is their evolution nearly independent on the whole cluster?.

We present a new method to identify subclusters in galaxy clusters which is based on coherent dynamical arguments. The algorithm chosen to construct this method is that of hierarchical clustering methods. It is based on the mathematical theory of cluster analysis (see, e.g., Anderberg 1973).

## 2. Hierarchical Clustering Method

The clustering analysis provides different mathematical methods to distribute a set of objects into a set of clusters. Among these methods, hierarchical clustering gives the clearest insight into the cluster structure and allows for a relatively easy identification of cluster members. This kind of methods can be formulated in terms of operational concepts (e.g., Schorr 1976) and it was first applied in the astronomy field to study how the galaxies in a sample are gathered in groups or clusters by Materne (1978), and later by Tully (1980, 1987) and Gourgoulhon et al. (1992).

### 2.1. General features of hierarchical methods

A hierarchical clustering method starts with a sample of $N$ objects (galaxies). The hierarchical relationship between these objects can then be obtained from the following procedure (see fig.1):

1) We initially consider that each object constitutes a group by itself, that is, we start by considering a collection of $N$ groups: $(\{G_1\}, \{G_2\}, ..., \{G_n\})$.

2) We define an affinity parameter $s$ (for instance: distances, inverse of forces,...), and an affinity or linkage function $S(s)$ (see below). We construct an array $S_{ij}$ giving the value of $S$ for each couple of groups $(G_i, G_j)$.

3) We then extract the two groups presenting the highest affinity: $\min(S_{ij})$. These two groups $G_k$, $G_l$ are then "merged" and replaced by a single group $G_{kl} = \{G_k, G_l\}$. So we are left with $N - 1$ groups: $(\{G_1\}, \{G_{kl}\}, ..., \{G_{n-1}\})$.

4) The merging procedure is repeated until only one cluster contains all the $N$ objects of the sample : $G_1$ is left.

5) The merging sequence can be easily visualized under the form of a hierarchical tree (h–tree). Such a graph shows from the smallest groups (or pairs) to the greatest structure including several groups (see Fig. 2).

### 2.2. Constructing a particular hierarchical method

The hierarchical method described above can be applied to several kinds of subjects. In order to construct in a

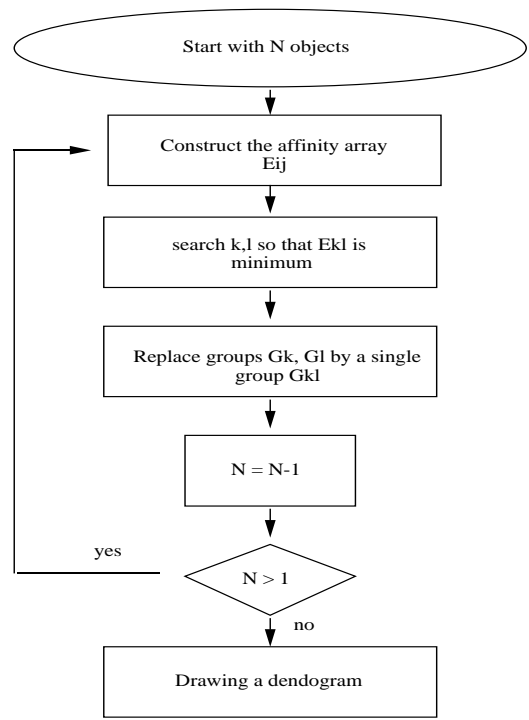

**Fig. 1.** Flow diagram corresponding to the hierarchical method

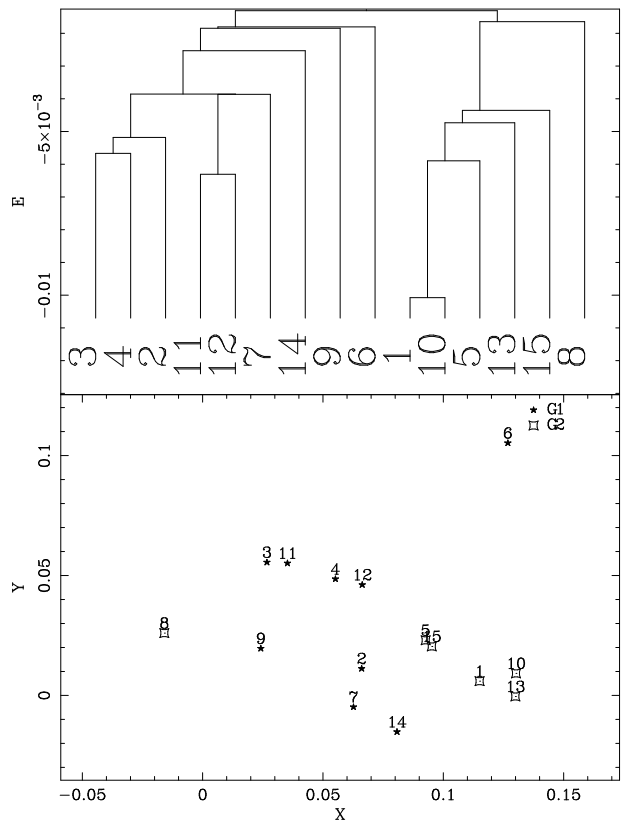

**Fig. 2.** Example of a hierarchical tree

coherent way a particular grouping method we need to specify:

1. What does a substructure mean?.
2. Which is the affinity parameter more closely associated to our substructure definition?.
3. Which linkage method gives the best results?

It is important to remark that the two last points cannot be separated from the first one. As a matter of fact, all the physical information specifying the nature of the particular problem under consideration is contained in the definition of the affinity parameter. Different choices of $S$ will therefore lead to different kinds of substructures in a sample. In the past, for instance, to extract groups and clusters from a sample of galaxies, Materne (1978) used the Ward criterion, which gives the increase, from two groups to their merged group, of some coordinate dispersion measure; Tully (1980) used a quantity which is related to the inverse of the force between two groups; and Gourgoulhon et al. (1992) used instead the inverse of the merged group density. However, these authors did not try to interpret the physical meaning of their resulting structures. It is then difficult to understand, from a dynamical point of view, the differences between the above methods and to decide which is the best suited one.

### 2.3. Substructure Definition

We then start by defining what a substructure means. Only provided with such a definition, we will be able to find and to test the best method to identify such a class of structures.

Observational experience just gives an intuitive idea of what a substructure is. An objective definition is however difficult to find. Simple definitions like "overdense regions" or "groups of galaxies with a mean spatial separation smaller than their local environment" have several inconveniences. For instance, such simple definitions would identify as a substructure a group of galaxies which, from a mere coincidence, are passing close together at a given instant.

In order to translate our intuitive idea to a more appropriate theoretical definition, we note that an isolated rich cluster has a membership which does not change through its dynamical evolution, except for some few escaping particles by collisional processes as ejection or evaporation. A system whose membership is considerably reduced in a relatively short time interval cannot be considered as a bound cluster. Such a situation is however much more complex for substructures within a galaxy cluster because they are not isolated. In principle, their membership could be modified through the system evolution by interactions with other substructures or by changes in the gravitational potential of the overall cluster. Nevertheless, the membership of a substructure must be much more "stable" than that of any random subset of galaxies without a particular dynamical signification. The most strongly bound substructures will have longer lifetimes than a weakly bound galaxy subgroup where the external influence is much more important. We have then decided to look for galaxy groups in a cluster with longer membership lifetimes than any random group. If they exist, we will call them "substructures".

### 2.4. The affinity parameter

As we have quoted above, since all the physical information is contained in the affinity parameter, a good choice of $s$ is crucial to obtain the greatest similarity between the obtained structures and those corresponding to our substructure definition. Since we are trying to find *gravitationally bound groups*, definitions of $s$ just based on the galaxy positions are not appropriate because, even if the force between two spatially close galaxies is rather strong, they could be not mutually bound if their relative velocity is high enough to overcome their gravitational attraction.

A measure of the boundness of a gravitational system is its *relative binding energy*.

$$E_{ij} = -G\frac{m_i m_j}{|r_i - r_j|} + \frac{1}{2}\mu_{ij}(v_i - v_j)^2 \qquad (1)$$

where

$$\mu_{ij} = m_i\, m_j/(m_i + m_j), \qquad (2)$$

Obviously, a negative $E_{ij}$ for two galaxies within a galaxy cluster does not necessarily imply that they are bound. Processes like close collisions or external gravitational influences could signify that they are unbound. Nevertheless, the more negative the relative binding energy is, the more strongly bound are two galaxies against the above quoted collisional or external processes. Such a feature is observed in several N-body simulations. For instance, Kandrup (1994) noted that "gravitational N-body simulations evidence that particles initially with high binding energy tend to end up with high binding energy, low with low, etc."

We then adopt such a quantity as our choice for $s$. In section 3.1 we will test, from N-body simulations, that an affinity parameter defined in terms of the relative binding energies is in fact appropriate to outline substructures defined by their longer membership lifetimes.

### 2.5. Linkage method

The cluster analysis theory provides three different linkage techniques to construct the affinity array $S_{ij}$ for a given choice of $s$:

1. *Single linkage:* each group is characterized by the longest $s$ value needed to connect any group member to some other member of that group.

$$S_{ij} = \min_{\substack{\mu \in G_i \\ \nu \in G_j}} (E_{\mu\nu}) \qquad (3)$$

2. *Complete linkage:* each group is characterized by the longest $s$ value needed to connect every group member to every other member.

$$S_{ij} = \max_{\substack{\mu \in G_i \\ \nu \in G_j}} (E_{\mu\nu}) \qquad (4)$$

3. *Centroid clustering:* each merged group is characterized by the $s$ value between its "parent" group means or center of mass properties.

$$S_{ij} = E_{\overline{G}_i \overline{G}_j} \qquad (5)$$

where $E_{\overline{G}_i \overline{G}_j}$ is the affinity (relative energy) between the mean of $i \in G_i$ and that of $j \in G_j$.

A comparative study of the main properties of these different linkage methods has been extensively performed in the literature (see, e.g., Anderberg 1973). The single-linkage method is one of the very few clustering techniques which can outline nonellipsoidal clusters, and the resulting h-tree is invariant to any monotonic transformation in the affinity parameter. It also has interesting connections with certain aspects of graph theory, as the problem of finding the minimum spanning tree. Its main disadvantage is that it can fail to distinguish two groups poorly separated in the $s$ space. Complete linkage is also invariant to monotonic transformations of the affinity measure. In contrast to the single-linkage method, it can only be interpreted in terms of the relationships within individual groups, while the differences between groups are not determined very reliably. Although it has some common aspects to a maximally connected subgraph, it has no special interpretation in graph theory terms. Finally, the centroid method and its variants has the great disadvantage that reversals can occur, i. e., the $S_{ij}$ value corresponding to the groups to be merged may rise and fall from stage to stage of the hierarchical "merging" sequence. When these reversals occur, the resulting h-tree has no meaningful interpretation.

In the next Section, we will also test from N–body simulations each of these three linkage methods. We will show that structures within a galaxy cluster are much better described by the single linkage clustering given by Eq.(3). We then adopt such a linkage technique to fully specify the particular hierarchical clustering method that we propose here.

## 3. Tests from N-Body Simulations

In this section, we will show the efficiency of our method by testing:

- The ability of a clustering method to extract those galaxy groups with the highest lifetimes. Different dynamical arguments seem to indicate that such a procedure is in fact appropriate to do it. However, it is necessary to test more rigorously this point not only by comparing with other possible choices of the affinity parameter $s$, but also by comparing the three main linkage techniques.

- The ability of the method applied on observational data (position on the sky and line-of-sight velocity i.e. 3-D–data) to recover the structures obtained using (6-D)–data. The efficiency of a clustering method must also be measured in terms of its applicability to real data, where some coordinates are unavailable. A method giving very different results when only the coordinates provided by observations are used, cannot be considered as useful in actual practice.

The first point will be analyzed by performing N-body simulations of the self-gravitating evolution of a galaxy system, while the second one will require a series of static numerical experiences.

In order to evaluate the stability of results when we compare, for a given system, two different evolutionary stages, or 6-D against 3-D configurations, we have implemented the following procedure. We start by applying the method under consideration to obtain the hierarchical trees (see Fig. 2) corresponding to each of the two configurations to be compared. For each of both trees, we construct an array defined so that the element $T_{ij}$ is the number of 'mergers' in the hierarchical sequence involving some group which contains simultaneously both $i$ and $j$ galaxies. The number of coincidences between the two trees (denoted by superindexes 1 and 2) is then

$$N_c = \sum_i \sum_{j \neq i} \min(T_{ij}^1, T_{ij}^2) \qquad (6)$$

while that of differences is

$$N_d = \sum_i \sum_{j \neq i} |T_{ij}^1 - T_{ij}^2| \qquad (7)$$

In order to quantify how different or similar are these two hierarchical trees, we then define the rate of instability or difference between them as

$$\mathcal{S} = \frac{N_d}{N_c + N_d} \qquad (8)$$

Obviously, the most stable method will lead to the smallest $\mathcal{S}$ values. Two exactly equal trees imply $\mathcal{S} = 0$, while those absolutely different lead to $\mathcal{S} = 1$.

### 3.1. Stability during the system evolution

We have considered in our N-body simulations a rather non-stationary system without very high density contrasts for substructures. Initial conditions were generated by considering 400 equal-mass particles with random positions within an uniform sphere and radial velocities (spherical collapse model). The self-gravitating evolution of that system was then simulated by using the N-body code developed by Alimi & Scholl (1993) on *Connection Machine*. At $t_1 = t_{col}/5$, some very slight condensations begin to be apparent and we have taken this configuration as our starting reference system. This configuration will then be compared to the later one at $t_2 = 2t_{col}/5$. Obviously, since substructures are not isolated and the overall gravitational potential is changing, the group memberships will vary somewhat from $t_1$ to $t_2$. We will now test the stability of structures defined by different methods during the system evolution.

### 3.1.1. The stability of linkage types

Fig.3a shows the resulting stability parameters $\mathcal{S}$ when each of the three linkage types (single, complete and centroid) are used together with the definition Eq.(1) for the affinity parameter. We see that the smallest evolutionary variations of substructures are obtained when a single linkage technique is used. Complete linkage leads instead to a $\mathcal{S}$ value greater by a factor of $\sim 2$ during the considered time interval, and centroid clustering leads to a even worse evolution stability.

This result shows that the single-linkage technique chosen (by us) in our hierarchical clustering method is indeed the best suited to obtain stable substructures. This result can be interpreted by the numerical artifacts produced when complete or centroid clusterings are used. As we have quoted in Section 2.5, centroid and complete clustering techniques are incapable of outlining non-ellipsoidal groups and they then introduce a spatial biasing in substructures. Moreover, if a centroid clustering is used, the value of $S_{ij}$ for two specific entities depends on the group sizes. This can result in a "snowball" effect where one of the first groups constructed by the hierarchical merging procedure starts to artificially accretate, due to its greater size, several single-component groups.

### 3.1.2. The stability of different hierarchical methods.

We have also tested the stability of our method (with a single linkage) as compared to other hierarchical methods proposed in the literature. Although some of these methods were originally proposed to analyze large-scale catalogs by using projected positions and, sometimes, also the line-of-sight velocities, their adaptation for substructures in a 6-D–simulated cluster is straightforward.

Fig.3b shows the stability parameter between $t_1$ and $t_2$ obtained when different hierarchical methods are used. We see again that our binding method is much more stable than other non-energy-based ones. In particular, the Ward criterion method proposed by Materne (1978) and also the density method by Gourgoulhon et al. (1992) are strongly unstable with the system evolution. Consequently, substructures defined by them do not correspond to bound subsystems in a cluster (except, presumably, if they are characterized by strong overdensities). An intermediate stability is found however for the inverse-of-force method by Tully (1980).

It is important to note that the good results obtained with our method are not simply due to the fact that it uses a larger amount of information (positions and velocities) than those just based on positions, like those involving forces or densities. The Ward criterion also uses velocities to measure the coordinate dispersion and, nevertheless, it leads to the worst results in terms of stability. We interpret this last result by the lack of dynamical coherence of such criterion. In the same way, one can expect that other dynamically non-coherent choices of the affinity parameter will also lead to unphysical results even if they use the same or much larger amounts of information.

### 3.2. Stability against projections

Each particle in a N-body simulation is characterized by its 6-D–coordinates $\{\mathbf{x}, \mathbf{v}\}$. However, astronomical observations only provide 3 coordinates $\{\alpha, \delta, v_z\}$. We will now test the stability of different methods when only 3-D coordinates are used, that is, we will test if the " observed" structures are reasonably similar to the "intrinsic" (6-D) ones.

The resulting stability parameters between $t_2^{6D}$ and $t_2^{3D}$ are shown in Fig.3c (similar results are found for $t_1$). We see that our method is the only one which does not imply considerably different results when projection biasing is present. The worst results are now obtained by Gourgoulhon's method, but Tully's and Materne's techniques also give very unstable results under projection effects. Concerning Tully's method, it must be noticed that, although it did not give very bad results when the evolution stability was examined, it implies however a rather poor similarity between observed and intrinsic substructures. This result clearly shows the efficiency of a hierarchical method based on relative energies as compared to other procedures. It is able to find relatively time-stable groups, and it also ensures a reasonable similarity between 3-D and 6-D structures.

This behavior can be understood as follows: We quoted above that the structure of the h-tree obtained by our method is invariant to any monotonic transformation in the affinity parameter. This is roughly the case when our affinity parameter is transformed by projection effects. The relative binding energy is in fact the sum of two terms, the potential energy and the kinetic energy. When computed by using only the observed coordinates, both terms are separately smaller (or more negative) than the intrinsic ones. Consequently, the "apparent" binding energy of any cluster galaxy will be smaller (more negative) than the 6-D one. However, since such effect is similar for all galaxies the relationships between them remain nearly unchanged. That is, two weakly bound galaxies close in the space of positions are usually also weakly bound under projection effects, as compared to the other ones, because their relative velocity holds rather high.

It must be however noticed that, when real data coming from observations are used, the situation could appear more complicated because interlopers are often present.

For instance, let us consider the case of a galaxy with "reasonable" position and line-of-sight velocity , but with a very hight transverse velocity. Any method will (probably) assign that galaxy to the "observed" cluster, while it does not actually belong to the 6-D-cluster. Such a situation, although possible, can be assumed very rare in real clusters.

The opposite case of a galaxy with a line-of-sight velocity much higher (or much smaller) than the other galaxies in the field, will be easily interpreted by the h-analysis as an interloper, or as belonging to a different system, even if their apparent positions are close. It is precisely the case of the cluster ABCG151, we will discuss in the next section.

## 4. Applications

The method we propose, based on the relative energy and a single linkage, is efficient and stable in the various senses described above. We will now use this tool to analyze substructures in galaxy clusters.

We transform angular distances between galaxies into distances in Mpc, by means of the cluster distance $D = \overline{v_z}/H_0$. $H_0 = 50$km s$^{-1}$ Mpc$^{-1}$ is the Hubble constant, and $\overline{v_z}$ is the mean redshift. Notice that when a clear partition is found after a first analysis, a further iterative study is performed by using the mean distances of subsystems. We have used in the computation of the energy, a mass-to-light ratio for the overall cluster of the same order than that obtained by the virial theorem. After several computations we found that the resulting hierarchical trees for real clusters depend rather weakly on these parameters, and they only indicate some differences if the mass-to-light ratio is over or underestimated by a factor of $\simeq 10^2$.

We have processed ABCG151 and ABCG2670 with the h–method.

The data for ABCG151 come from the **ESO**–**K**eyprogram on **N**earby **A**bell **C**lusters of galaxies (ENACS Mazure et al 1991). The sample is described in den Hartog (1995). Notice that our data include all galaxies in the direction of ABCG151, that is, we have not used redshift criteria to eliminate interloopers, foreground or background groups (see Fig. 4). This will allow us to show how such separation is performed by the h–method.

Concerning ABCG2670, we have used the data given by Sharples et al. (1988). In their paper, these authors provide us with a subsample of the galaxies belonging to the cluster (based on the redshift criteria), and limited to magnitude $b_j = 19$. We used only those galaxies located in a circle of radius $r = 1200$arcsec centered on the cD. This sample contains 48 galaxies. The velocity histogram is displayed in Fig. 6.

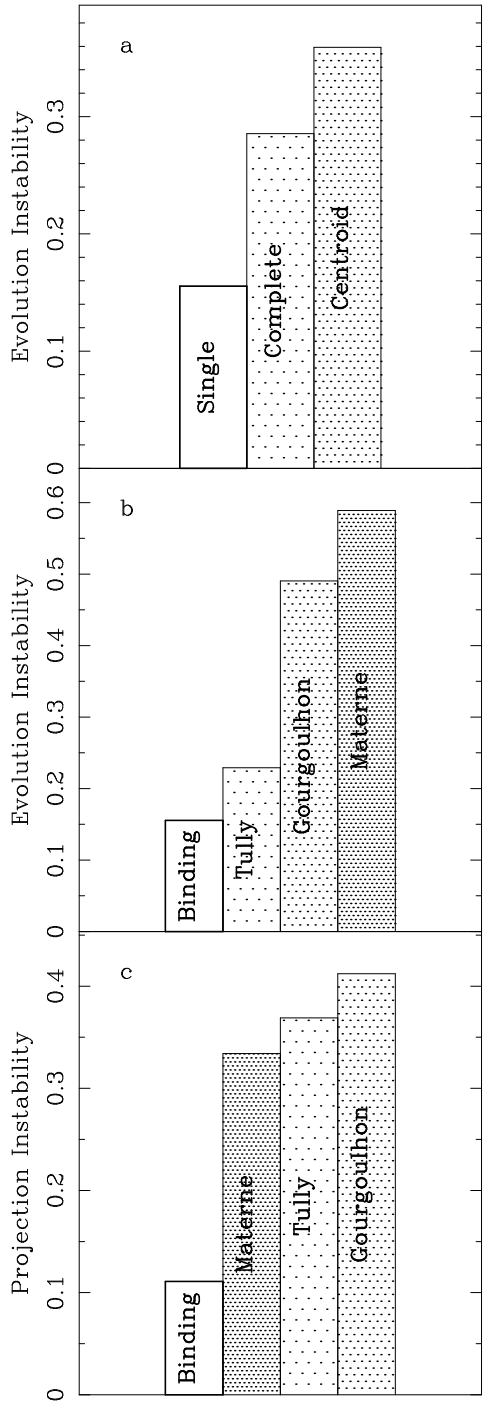

**Fig. 3.** Instability of hierarchical trees against dynamical evolution (a and b) and projection effects (c) for different linkage types (a) and different hierarchical methods (b and c)

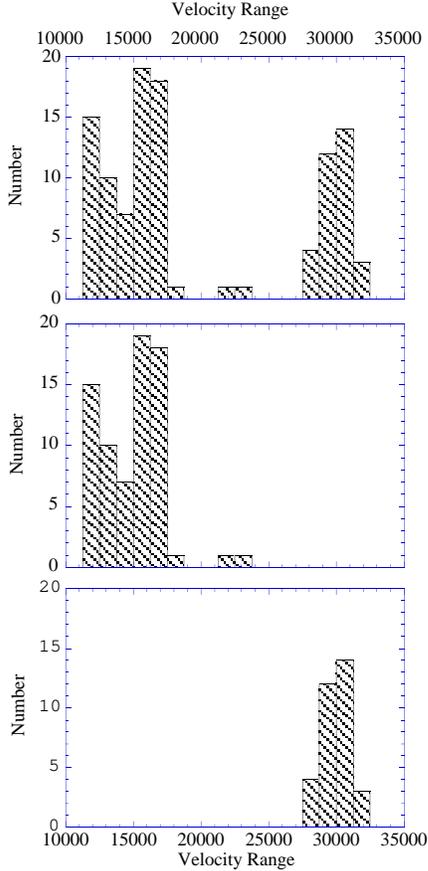

**Fig. 4.** Velocity histogram. Velocities are in km s$^{-1}$. The whole field of ABCG151, G1 and G2 (see text)

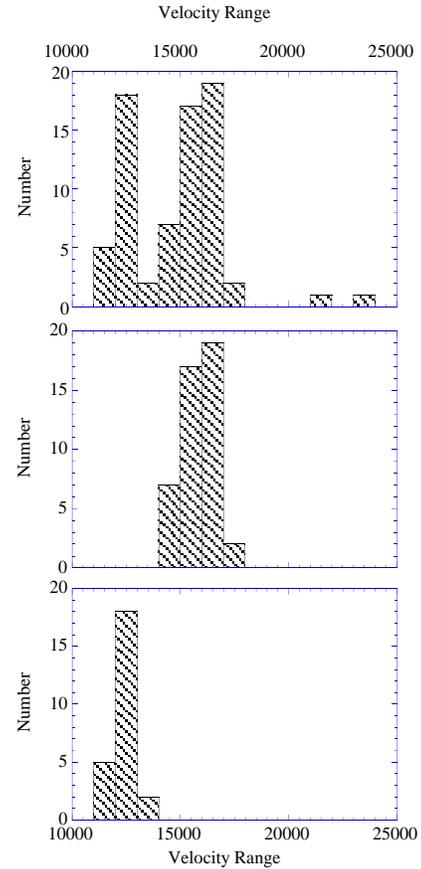

**Fig. 5.** Velocity histogram. Velocities are in km s$^{-1}$. ABCG151– G1, G(1,1) and G(1,2) (see text)

### 4.1. The cluster ABCG151

We can clearly see on the h–tree shown in Fig. 7 that ABCG151 is constituted of two main clusters, G1 and G2. The method give explicitly those galaxies belonging to each of these clusters. G1 and G2 are superimposed on the sky, but they are separated in the "h-tree" (fig.7) as well as in the velocity histogram (fig.4). The mean velocity of G1 is $V = 14896$km s$^{-1}$, while that of G2 is $V = 29900$km s$^{-1}$. If their velocities were only due to the Hubble expansion, the distance between the two clusters would be $\approx 300$Mpc

G1 is subdivided into two subclusters G(1,1) and G(1,2). These two subclusters have nearly similar mean velocities as compared to the G2 velocity (see table 1). Clearly separated in the sky (Fig.7),they perhaps actually form a pair of clusters. It is interesting to notice that these two clusters have velocity dispersions usual for clusters or large groups (see also Fig.5). The analyses performed by the ENACS group, using fixed–gap and robust estimator for the velocity dispersion also lead to three clusters with the characteristics indicated in table 1(bottom).

Finally, G2 has no clear subclustering. The h–tree nevertheless shows a well bound *pair of pairs* in the deepest energy levels with some other neighbour galaxies forming a central dynamical core. A similar central structure will be found in the case of ABCG2670 discussed below (Sect. 4.2).

As we have explained above, we have also performed a h–analysis on G1 and G2 separately, but the resulting h–trees indicate minor changes only.

### 4.2. The cluster ABCG2670:

The first result we emphasize is that the h–tree does not reveal subclustering in ABCG2670 (Fig.8).

However more can be said: the deepest energy levels in the ABCG2670 h-tree show a well bound *pair of pairs* formed by galaxies (1, 19) and (6,8). It is interesting to remark that galaxy 1, the cD one, is located at the bottom of the energy well, and galaxy 19 is very close to it. The same behavior occurs for the pair (6,8) which is also located very close to the cluster center. These two pairs, together with galaxies 44, 27, and 31 define a bound central group (denoted"b-c-g") in the deepest part of the energy

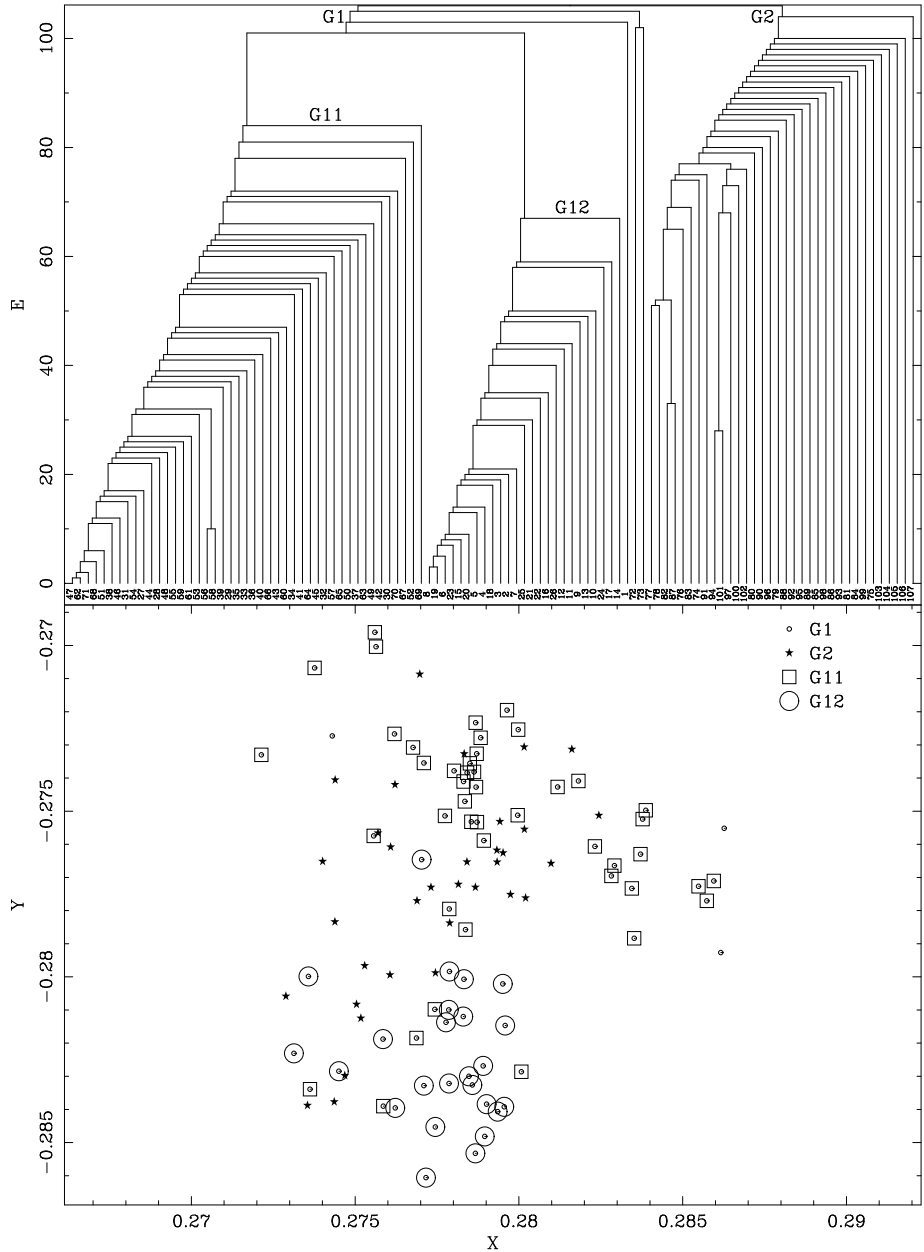

**Fig. 7.** A151 substructure analysis. a – the structural tree. b – the map

levels. Again notice the similarity of central structures in ABCG2670 and ABCG151–G2.

The velocity characteristics of ABCG27670 are displayed in table 2.

The result we found is very different from that given by C. Bird (1994) which is based on the analysis of the galaxy velocities and the hypothesis that subgroups have a Gaussian velocity distribution . However, this kind of hypotheses is questionable: we may in fact notice that the total luminosity of the cD galaxy is almost twice the total luminosity of all the remaining "b-c-g"–galaxies. This feature suggests that these galaxies are rather a satellite system around gal.1 instead of a virialized group with a Gaussian distribution of velocities.

## 5. Comments on the hierarchical method

Motivated by questions like: how can subclusters be defined from a dynamical point of view?; are they just transient phenomena?, is their internal dynamics nearly independent on the cluster where they are included?; we have developed a new method to identify substructures in galaxy clusters.

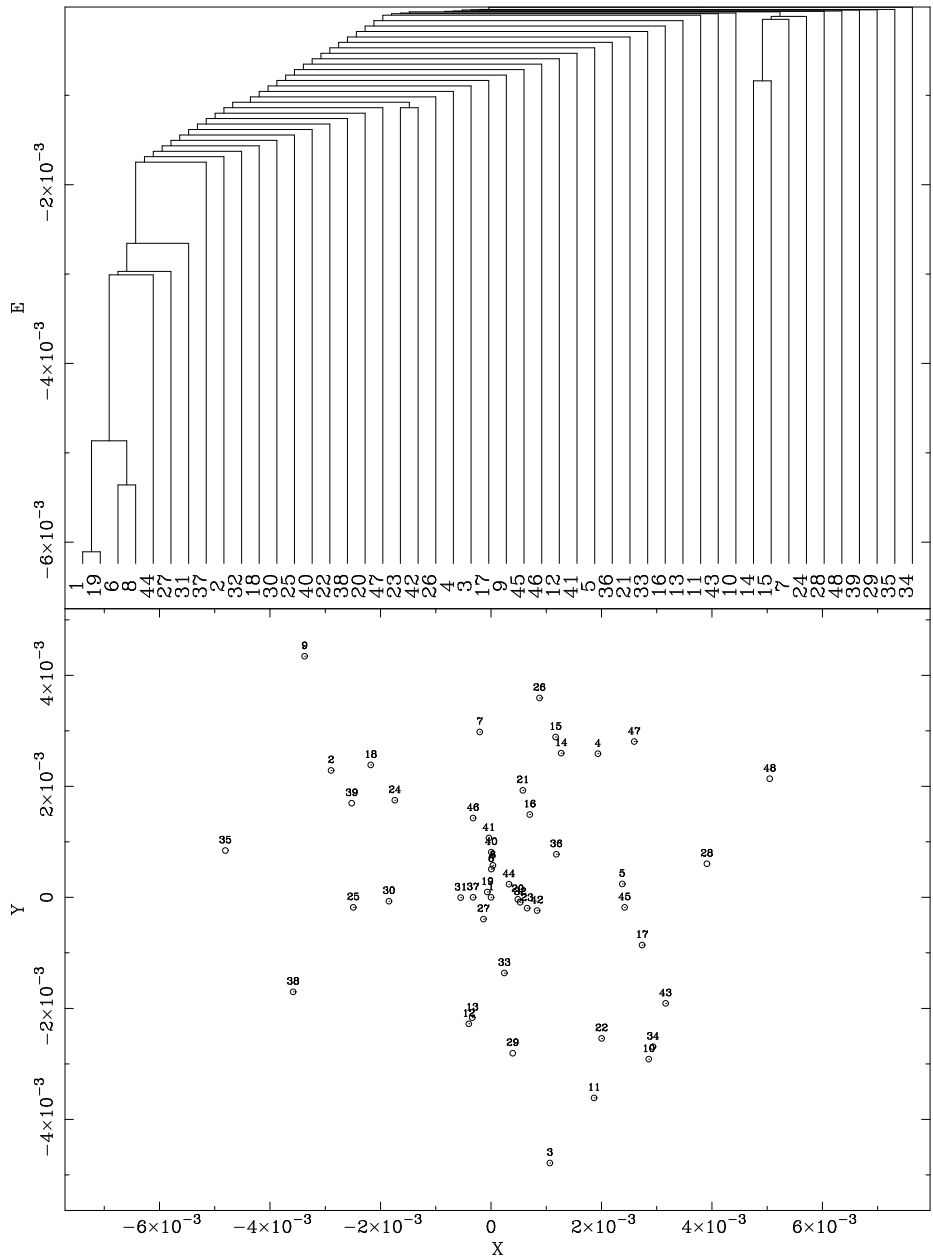

**Fig. 8.** A2670 substructure analysis. a – the structural tree. b – the map

Since observational experience just provides us with an intuitive idea of what a substructure is, an objective answer to the first of the above questions is difficult to find. We have translated this intuitive idea to a theoretical definition by looking for subsamples with longer lifetimes than any random subset of galaxies in a cluster. If they exist, we call them "substructures". In order to implement such a definition, we have used a hierarchical clustering algorithm to determine the relationship between the cluster galaxies according to their relative binding energies. Obviously, a substructure definition different from what we propose could require a different working parameter.

Since overdense regions in real clusters usually satisfy our above definition, the substructures defined by our method are not far from the intuitive idea of such kind of entities. Moreover, the inclusion of relative velocities in the kinetic term of binding energies (in addition to the spatial information contained in the potential term), results in a reasonable similarity between the observed and the intrinsic substructures. In other words, the h-method is able to differentiate a cluster group from other groups or isolated galaxies just superimposed on the same region of the sky, to include galaxies dynamically bound to a group but with orbits or positions far from the region clearly covered by that group, and to identify subclusters not characterized by a remarkable overdensity in a cluster. All these features, as well as the evolution stability of structures found by our method, have been tested by means of numerical experiments and N-body simulations.

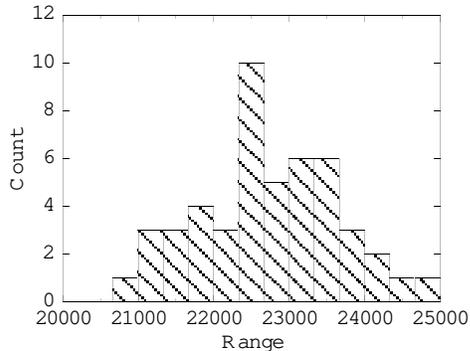

**Fig. 6.** Velocity histogram of ABCG2670. Velocities are in km s$^{-1}$.

|        | numbers | mean  | median | Vel. Disp. |
|--------|---------|-------|--------|------------|
| field  | 105     | 19612 | 16310  |            |
| G1     | 72      | 14896 | 15515  | 2230       |
| G2     | 33      | 29900 | 30030  | 846        |
| G(1,1) | 45      | 15955 | 15986  | 748        |
| G(1,2) | 25      | 12389 | 12287  | 390        |
|        | 40      | 29910 |        | 857        |
|        | 62      | 15960 |        | 667        |
|        | 29      | 12300 |        | 395        |

**Table 1.** Velocity characteristics of ABCG151. Top – from the hierarchical method – Field means: all the items in the field of the so-called "cluster". G1, G2, G(1,1) and G(1,2). Bottom – ENACS analysis – Clusters are separated in the velocity space using a fixed gap. Velocities are in km s$^{-1}$.

|         | numbers | mean  | median | Vel. Disp. |
|---------|---------|-------|--------|------------|
| A2670   | 48      | 22744 | 22646  | 840        |
| cD      |         | 23280 |        |            |
| "b-c-g" | 7       | 22962 | 22856  | 571        |

**Table 2.** Velocity characteristics of ABCG2670.b–g for the central bounded group, see text. Velocities are in km s$^{-1}$.

Concerning the transitory character of substructures, the results obtained in Sec.3.1 from N-body simulations show that their lifetime strongly depends on how they are defined and on the algorithm used to find them. Other previous methods as those referred to in Section 3.1, define substructures which could just be transient entities with rather short lifetimes. Our results show however that it is possible to define subclusters in such a way that they are much more stable throughout the cluster evolution. Groups located in the h-tree at the deepest energy levels will have *a priori* the longest lifetimes. Notice by the way that, since our method is able to extract substructures subsisting during long times, it appears specially useful for those works looking for substructures as a discriminant tool between various formation scenarios.

Substructures are often considered in the literature as systems isolated from the whole cluster. We would like however to emphasize some aspects concerning this point:

From a rigorous point of view, they can never be considered as isolated because they are embedded in the cluster. Their dynamics are then influenced by other substructures and also by the overall cluster. Nevertheless, by looking at the h–trees obtained for clusters like ABCG151 and AGCG2670, we see that some subgroups are characterized by a strong increase in their binding energy. This is for instance the case of galaxies (1,19,6,8) in the h–tree shown in Fig.8. A priori, it can be expected that such a class of subsystems tend to be much more isolated, or independent, than other groups characterized by weaker increases of their energy levels. Our h–tree technique then provides an idea of how good is the approximation of "isolated" for a given galaxy group.

Much work is still needed to improve the h–method. For instance, by including an estimation of confidence levels of the obtained substructures; by taking into account the presence of a background non-luminous component (X-ray emitting gas and dark matter) dominating the gravitational potential. Although reasonable results could be expected if light traces the mass, a rigorous study of this problem is needed, specially in the opposite case with a considerable luminous-dark matter segregation. These improvements are in progress.

*Remarks:* Information on the h–program can be provided by e-mail at: serna@gin.obspm.fr

*Acknowledgements.* We thank V. G. Gurzadyan for useful discussions. We are specially indebted to J.-M. Alimi for very valuable comments on this work and for allowing us to use his N-body code. We thank F. Durret for useful comments and carefull reading of the manuscript.

A. Serna thanks the Ministerio de Educación y Ciencia, Spain, for a post-doctoral fellowship. The N-body computations were carried out on the CM-5 of the Institute de Physique de Globe, Paris.

## References

Alimi, J.-M., & Scholl, H. 1993, Int. J. Mod. Phys. C 4, 197
Anderberg, M. R. 1973, *Cluster analysis for applications* (London: Academic Press)
Baier, F. W., 1977, Astr. Nach., 298, 151.
Bird, C. M. 1994, ApJ, 422, 480.

Bird, C. M. 1995, AJ (in press).
den Hartog,R. 1995, *The Dynamics of Rich galaxy Clusters* Thesis, Sterrewacht Leiden, Nederland
Jones,C., Forman,W., 1992, in *Clusters and superclusters of galaxies*,ed. Fabian,A.C.,(Dordrecht;Kluwer),p. 49
Geller, M. J., & Beers, T. C., 1982, PASP 94, 421.
Gourgoulhon, E., Chamaraux, P., & Fouqué, P. 1992, A&A 255, 69.
Grebenev, S.A., Forman,W., Jones,C. & Murray, S., 1995 Harvard-Smithsonian CFA preprint N. 4010, To appear in ApJ, April 20.
Kandrup,H.E.1994 in *the Seventh Marcel Grossmann Meeting*
Kauffmann, G., & White, S. D. M. 1993, MNRAS, 261, 921.
Lacey, C. G., & Cole, S. 1993, MNRAS 262, 627.
Materne, J. 1978, A&A, 63, 401
Mazure, A. et al. 1991, in *The Distribution of Matter in the Universe*, eds G.A. Mamon & D. Gerbal ( Paris, Observatoire de Paris) p.79
Merritt, D. 1985, ApJ, 289, 18.
Richstone, D., Loeb, A., & Turner, E. L. 1992, ApJ, 393, 477.
Salvador–Sole, E., Sanroma, M., González–Casado,, G. 1993, ApJ, 402,398
Serna, A., Alimi, J.-M., & Scholl, H. 1994, ApJ, 427, 574.
Schorr, B. 1976, CERN Data Handling Division, Internal Report 76-3.
Sharples, M. C., Ellis, R. S., & Gray, P. M. 1988, MNRAS, 231, 479
Slezak,E., Durret, F. & Gerbal,D. 1994, A.J., 108(6),1996
Tremaine, S. 1990, in *Dynamics and Interactions of galaxies*, ed. R. Wielen, (Berlin: Springer-Verlag), p. 394.
Tully, R. B. 1980, ApJ, 237, 390.
Tully, R. B. 1987, ApJ, 321, 280.
West, M.J. 1994a, MNRAS 268,797.
West, M.J. 1994b, in *Clusters of Galaxies* (XIVthMoriond Astrophysics Meetings), eds Durret,F., Mazure,A. & Tran Thanh Van,J., (Edtions Frontieres, France)